\documentclass{article}

    \PassOptionsToPackage{numbers, compress}{natbib}

    \usepackage[preprint]{neurips_2023}

\usepackage[utf8]{inputenc} 
\usepackage[T1]{fontenc}    
\usepackage{hyperref}       
\usepackage{url}            
\usepackage{booktabs}       
\usepackage{amsfonts}       
\usepackage{nicefrac}       
\usepackage{microtype}      
\usepackage{xcolor}         
\usepackage{natbib}
\usepackage{graphicx}
\usepackage{float}
\usepackage{subcaption} 

\title{A Comparative Study of CNN, ResNet, and Vision Transformers for Multi-Classification of Chest Diseases \thanks{Code and Data are available: https://github.com/Aviral-03/ViT-Chest-Xray}}

\author{
    Kaushik Murali \\
    University of Toronto \\
    \texttt{kaushik.murali@mail.utoronto.ca} \\
    \And
    Isha Surani \\
    University of Toronto \\
    \texttt{isha.surani@mail.utoronto.ca} \\
    \And
    Aviral Bhardwaj \\
    University of Toronto \\
    \texttt{aviral.bhardwaj@mail.utoronto.ca} \\
    \And
    Ananya Jain \\
    University of Toronto \\
    \texttt{ananya.jain@mail.utoronto.ca} \\
}

\begin{document}

\maketitle

\begin{abstract}
Large language models, notably utilizing Transformer architectures, have emerged as powerful tools due to their scalability and ability to process large amounts of data. \citet{dosovitskiy2021image} expanded this architecture to introduce Vision Transformers (ViT), extending its applicability to image processing tasks. Motivated by this advancement, we fine-tuned two variants of ViT models, one pre-trained on ImageNet \cite{imagenet} and another trained from scratch, using the NIH Chest X-ray dataset \cite{wang2017chestxray8} containing over 100,000 frontal-view X-ray images. Our study evaluates the performance of these models in the multi-label classification of 14 distinct diseases, while using Convolutional Neural Networks (CNNs) and ResNet architectures \cite{resnet} as baseline models for comparison. Through rigorous assessment based on accuracy metrics, we identify that the pre-trained ViT model surpasses CNNs and ResNet in this multilabel classification task, highlighting its potential for accurate diagnosis of various lung conditions from chest X-ray images.
\end{abstract}
\section{Introduction}

Detecting diseases early and accurately is important for the treatment and improvement of patient outcomes. Although chest X-ray imaging is a relatively low cost tool for diagnosis, radiologists are needed to analyze these images. However the limited access to radiologists in many areas, and the variability between radiologists can be a problem \cite{kermany2018identifying}. Machine learning provides a promising solution to increase detection accuracy and make medical image analysis available in areas without access to radiologist services. This approach may also lead to a more nuanced and precise classification of diseases due to the ability of models to learn relevant features from medical images \cite{sarvamangala2022convolutional}. Machine learning algorithms also have the ability to process large amounts of medical images leading to a more timely diagnosis and potentially the identification of issues overlooked by radiologists \cite{liu2023recent}. 

Recently, there have been significant breakthroughs via the application of deep learning techniques. Among these techniques, Convolutional Neural Networks (CNNs), Residual Networks (ResNet), and Vision Transformers have proven to be important in improving the precision and efficiency of these tasks. CNNS \cite{cnn} are a good choice for tasks like image segmentation due to their ability to recognize patterns in data. ResNet \cite{resnet} uses skip connections to address the vanishing gradient issue present in neural networks. Vision Transformers \cite{dosovitskiy2021image} have adapted the transformer architecture used on text for image classification, leading to impressive performance with reduced computation. In our project, we will conduct a comparative study of these three architectures in the multi-classification of chest cancerous cells, to identify the most effective approach for the multi-classification of chest cancerous cells.

\section{Related Work}

\textbf{\citet{kermany2018identifying}} focused on presenting the applications of deep learning models, particularly CNNs, to precisely detecting pneumonia in chest X-ray images. This study offers insights into how CNNs can be applied practically to the medical field in the context of medical imaging. The researchers highlighted the strengths of CNNs in capturing detailed textural and morphological patterns that are crucial to identifying pneumonia signs. Their findings emphasize the practical benefits of employing CNNs in the medical field, while also resolving obstacles such as the variability in image quality and the model's sensitivity to overfitting.

\textbf{\citet{liu2023recent}} highlighted how transformer models, through their attention mechanisms, are proficient at processing medical images for tasks such as disease detection, classification, and segmentation. \citet{liu2023recent}, reported that transformers achieve higher accuracy and efficiency, particularly in handling images with complex pathological features due to their ability to focus on relevant segments of an image without being constrained by the spatial hierarchies that typical CNNs rely on. The study not only offers us insights into how transformers have great potential in medical image analysis but also guides us in tackling obstacles we might face such as computational demands.

\section{Approach}
In our study, we focused on three architectures Convolutional Neural Networks (CNNs), Residual Networks (ResNet), and Vision Transformers.

\begin{figure}[htbp]
  \centering
  \begin{subfigure}[b]{0.4\textwidth}
    \includegraphics[width=\textwidth]{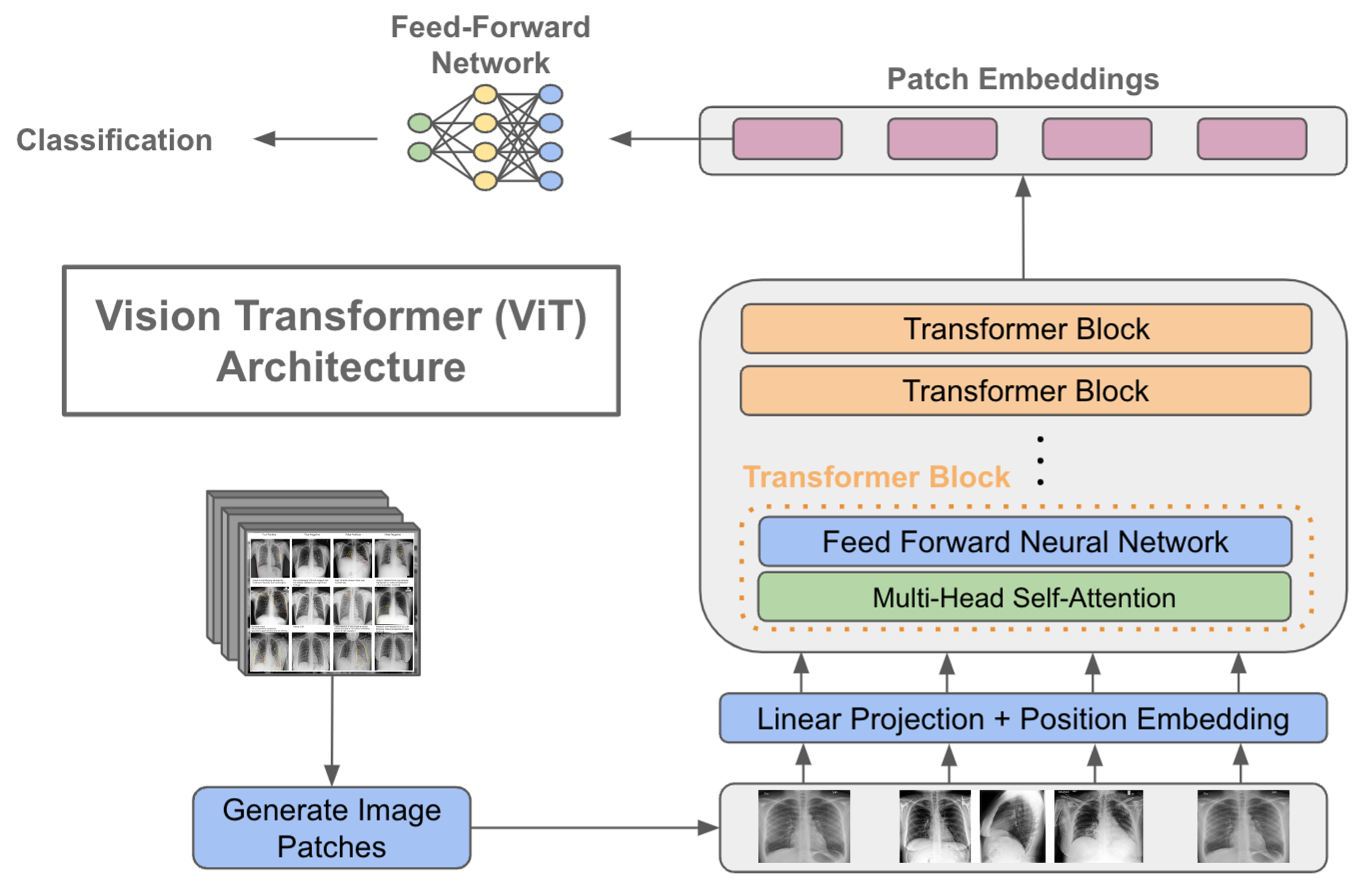}
    \caption{Sample ViT for Chest X-Ray \cite{transformers}}
    \label{ViT}
  \end{subfigure}
  \hfill
  \begin{subfigure}[b]{0.5\textwidth}
    \includegraphics[width=\textwidth]{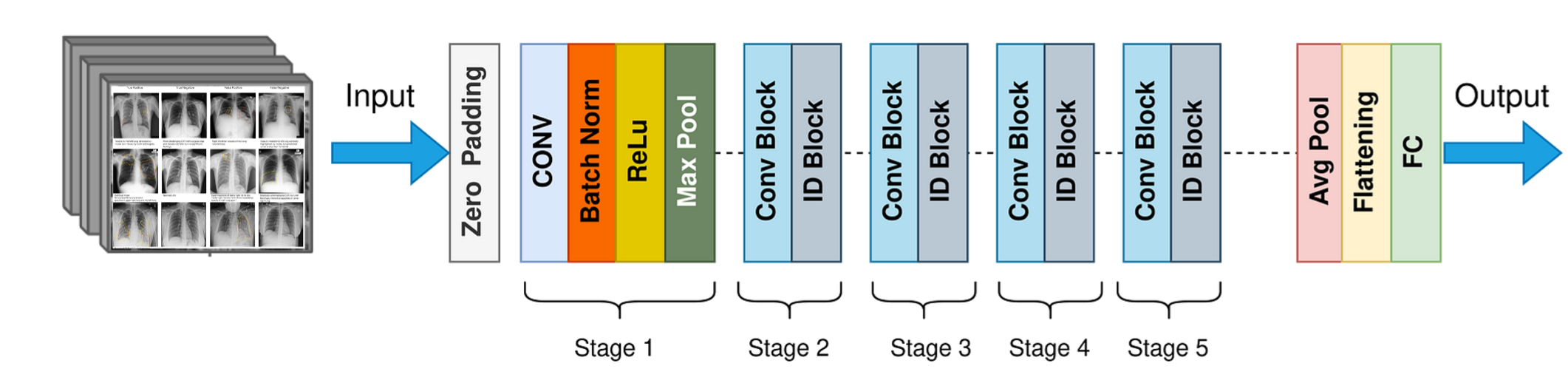}
    \caption{ResNet50 Architecture (Sample Implementation)}
    \label{ResNet}
  \end{subfigure}
  \caption{ViT, and ResNet Model Illustration}
\end{figure}

\subsection{CNN}
\textbf{Convolutional Neural Networks (CNNs)} \cite{cnn} are our baseline model, selected for their proven capability in medical image analysis \cite{sarvamangala2022convolutional}. Employing the Keras Sequential model, the CNN is structured with an input layer for 224x224 pixel RGB images, followed by a convolutional layer with 32 (3x3) filters and ReLU activation to detect basic features, a max pooling (2x2) layer to reduce spatial dimensions and improve invariance, another convolutional layer with 64 (3x3) filters for higher-level feature extraction, a second max pooling layer to further reduce dimensions, a flatten layer to convert maps to vectors, a dense layer with 512 ReLU units for pattern learning, and a final dense layer with ‘num\_classes’ sigmoid units for classification.

\subsection{Residual Networks (ResNet) \ref{ResNet}}
We will use \textbf{Residual Networks (ResNet)} \cite{resnet}, which use skip connections to train very deep networks, in order to solve the vanishing gradient problem that might arise with conventional deep CNNs. \\
The model structure starts with an input layer, which accepts images of size $224 \times 224$ with 3 colour channels. This is followed by a convolutional layer of $7 \times 7$ filters (stride 2) which extracts important low-level features from the images, batch normalization to normalize the activations, ReLU to introduce non-linearity, and max pooling (3x3, stride 2) to further reduce dimensions and introduce translation invariance. 

Subsequently, there are ResNet building blocks which each consist of two $3 \times 3$ convolutional layers, each followed by batch normalization and ReLU activation, and skip connections which are used to help prevent the vanishing gradient problem. There are 3 blocks of 64 filters, 4 blocks of 128 filters, 6 blocks of 256 filters, and 3 blocks of 512 filters. This is followed by pooling and a dense layer with sigmoid activation for multi-label classification.

\subsection{Vision Transformers (ViTs)\ref{ViT}}
\begin{figure}
    \centering
    \begin{subfigure}[b]{0.45\textwidth}
        \centering
        \includegraphics[width=\textwidth]{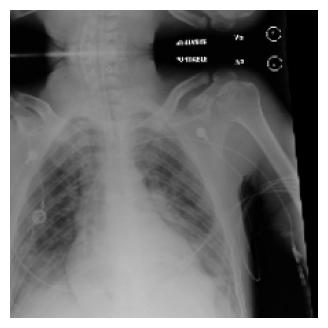}
        \caption{Input Chest X-Ray Image $224 \times 224$}
        \label{fig:input_image}
    \end{subfigure}
    \hfill
    \begin{subfigure}[b]{0.45\textwidth}
        \centering
        \includegraphics[width=\textwidth]{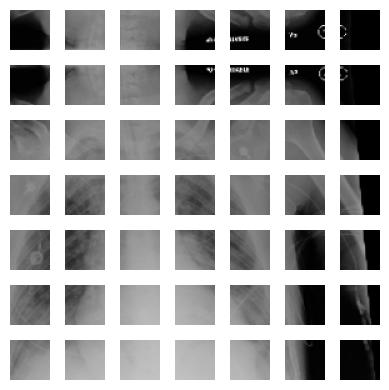}
        \caption{Image Patches $32 \times 32$}
        \label{fig:patches}
    \end{subfigure}
    \caption{ViT-v1/32 and ViT-v2/32 Input}
\end{figure}

Vision Transformers (ViTs) \cite{dosovitskiy2021image} utilize transformer architecture to advance image classification, processing $ 224 \times 224$ pixel images \ref{fig:input_image} into ($ 32 \times 32$) \ref{fig:patches} patches considered as tokens, which are then projected into a higher dimension with positional embedding. Our vision transformer models include multiple transformer blocks with layer normalization to stabilize input features, multi-head attention for segment focus, skip connections to aid gradient flow, a multi-layer perceptron (MLP) with GELU activation sequence to enhance learning, and a final layer normalization before dropout-enhanced MLP outputs through a sigmoid activation layer for multi-label classification, effectively leveraging the transformer's strengths for complex spatial hierarchies in medical imaging. 

\section{Experiment}
\subsection{Dataset}
To evaluate the performance of our model architectures, we utilized two freely available datasets: the \textbf{NIH Chest X-ray dataset} \cite{wang2017chestxray8} comprising 112,120 X-ray images with disease labels from 30,805 unique patients, and a Random Sample of the NIH Chest X-ray Dataset, containing 5,606 X-ray images. Both datasets involved multi-class classification across 15 classes, each representing different disease labels.

To optimize performance for specific tasks, we performed manual hyper-hyperparameter tuning for the ViT-v1/32 and ViT-v2/32 models through trial and error. This tuning was performed using the Random Sample of the NIH Chest X-ray data set. However, our model training was conducted on a subset of 85,000 images from this Random Sample Dataset. We observed  a faster convergence to optimal outputs within this subset.

We compare the performance of CNN, ResNET, ViT-v1/32, ViT-v2/32 model, ViT-ResNet/16 on our NIH Chest X-ray dataset, with 85000 images. Table \ref{tab:summary}, shows the summary of the transformers architecture that we implemented.

\subsection{Models}
Different designs of ViT models were used in our implementation. ViT models, name ViT-v1/32 which is our base variation with $32 \times 32$ input patch size, ViT-v2/32 variant $32 \times 32$ input patch size, and finally ViT-ResNet/16 variant with $16 \times 16$ input patch size which was pre-trained on ImageNet-21k \cite{imagenet}, a form of self-supervised learning. For comparison of the transformer, we also trained CNN models and ResNet models.

For ViT-v2/32 we explored several optimizers and observed that AdamW and SGD yielded similar results, however, the SGD training time was 10\% faster. In both models, we used `ReduceLROnPlateau` optimizer.

\begin{table}[h]
\caption{Summary of ViT Models: Hyper-parameter Choices}
\label{tab:summary}
\begin{tabular}{|c|c|c|c|c|}
\hline
Model & Optimizer & Metric & Loss & Pre-trained \\
\hline
ViT-v1/32 & AdamW & AUC & Binary Cross Entropy & - \\
ViT-v2/32 & SGD & AUC & Binary Cross Entropy & - \\
ViT-ResNet/16  & Adam & - & Binary Cross Entropy with Logits & ImageNet-21k \\
\hline
\end{tabular}
\end{table}

We also performed various data augmentations on both datasets. For the Chest X-ray dataset, we applied resizing, random horizontal flip, and random rotation.

\subsection{Evaluation Metrics}
We have plotted ROC and AUC Curve \ref{fig:ROC_ViT} and compared the performace of model of the NIH Chest X-ray dataset.

The Receiver Operating Characteristic (ROC) curves provide valuable insights into the performance of our multi-label classification model across the various lung disease classes present in the chest X-ray dataset. Each curve represents a specific disease class, with the Area Under the Curve (AUC) quantifying the model's ability to discriminate between positive and negative instances for that class.

\subsection{Results}

The performance of the trained models was evaluated using various quantitive indicators, which consisted of Test Accuracy, Validation Accuracy, Train Accuracy over 10 Epochs on the \textbf{NIH Chest X-ray dataset} \cite{wang2017chestxray8}. The results are shown in Table II, showing compelling results where ResNet consistently outperformed for ViT-v1/32 and ViT-v2/32 as it gave sub 93\% accuracy across all evaluation metrics.

Following on these results, we have therefore introduced ViT-ResNet. Which used ViT state-of-the-art attention mechanism, and ResNet capabilities of extracting features. This yeiled similar results to ResNet, however, with 10\% faster training time and much better ROC curve. In summary, 0.2 is the best validation loss that can be achieved based on the multiple experiments that we conducted.

\begin{table}[h]
\centering
\caption{Validation Accuracy of Models}
\label{tab:accuracy}
\begin{tabular}{|c|c|c|c|c|}
\hline
Model & Test Accuracy (\%) & Train Accuracy (\%) & Validation Accuracy (\%) & AUC \\
\hline
CNN & 91 & 92.62 & 92.68 & 0.82 \\
ResNet & 93 & 93.38 & 93.34 & 0.86 \\
ViT-v1 & 92.63 & 92.7 & 92.89 & 0.86 \\
ViT-v2 & 92.83 & 92.94 & 92.95 & 0.84 \\
ViT-ResNet & 93.9 & 93.02 & 94.07 & 0.85 \\
\hline
\multicolumn{5}{l}{Note: Train and Validation refers to Top Accuracy. Training curves are included \ref{appendix}} \\
\end{tabular}
\end{table}

\begin{figure}
    \centering
    \begin{subfigure}[b]{0.45\textwidth}
        \centering
        \includegraphics[width=\textwidth]{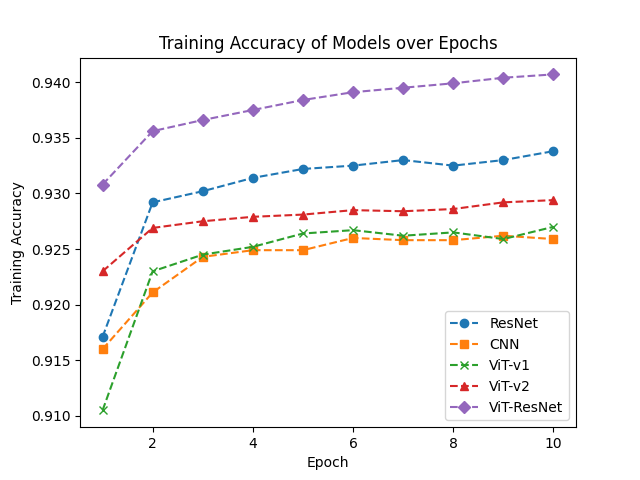}
        \caption{Training Accuracy of Models}
        \label{fig:ROC_ViT}
    \end{subfigure}
    \hfill
    \begin{subfigure}[b]{0.45\textwidth}
        \centering
        \includegraphics[width=\textwidth]{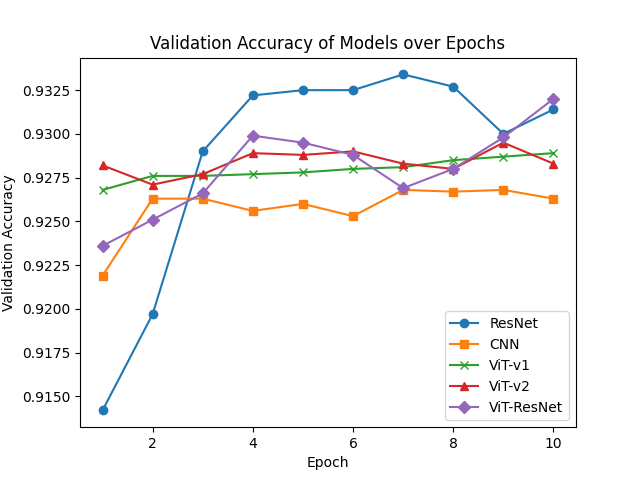}
        \caption{Validation Accuracy of Models} 
        \label{fig:ROC_CNN}
    \end{subfigure}
    \caption{Training and Validation Accuracy for various Model on NIH Chest X-Ray Images}
\end{figure}

\section{Discussion}

Our research focused on a comparative study that evaluated the performance of various Vision Transformers against CNN and ResNet models. As indicated in Table II \ref{tab:accuracy} and Figure 3, the transformer architectures consistently outperformed CNN models on all metrics, only achieving the test accuracy 91\%. This is because for large datasets CNN is unable to generalize and extract relevant features.

In particular, both ViT-v1 and ViT-v2 were surpassed by ResNet in performance. Several factors may contribute to this outcome. Firstly, the ViT models may not have been pre-trained on sufficiently large datasets, which could limit their ability to capture diverse and complex features present in the X-ray images. Furthermore, these models might not have been optimized specifically for feature extraction tasks relevant to medical image analysis, potentially leading to sub-optimal performance compared to ResNet.

However, it is noteworthy that comparable performance was observed for ViT-ResNet, where the transformer model was pre-trained on the ImageNet-21k  dataset. This suggests that pre-training on a larger and more diverse dataset can significantly enhance the performance of transformer architectures, aligning them more closely with established CNN models such as ResNet, with test accuracy as high as 94\%.

\subsection{AUC and ROC Curve}

\begin{figure}
    \centering
    \begin{subfigure}[b]{0.45\textwidth}
        \centering
        \includegraphics[width=\textwidth]{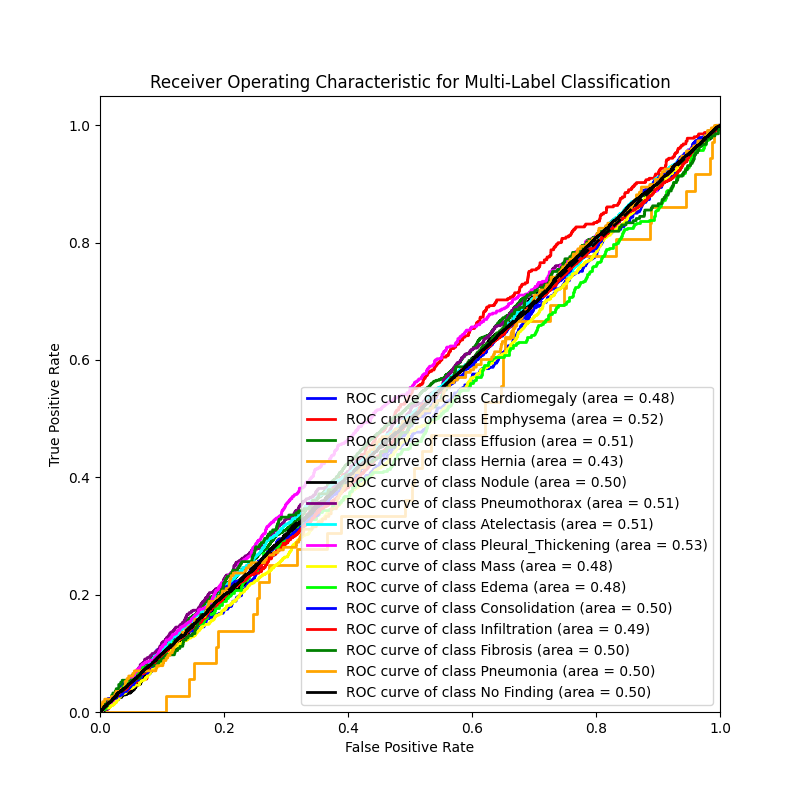}
        \caption{ROC curve of ViT (Larger Dataset)}
        \label{fig:ROC_ViT}
    \end{subfigure}
    \hfill
    \begin{subfigure}[b]{0.45\textwidth}
        \centering
        \includegraphics[width=\textwidth]{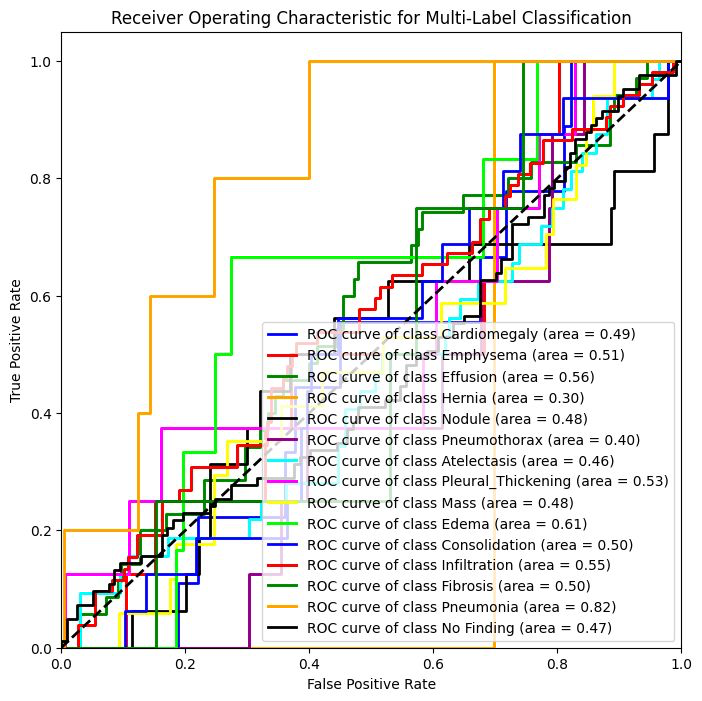}
        \caption{ROC curve of CNN (Smaller Dataset)}
        \label{fig:ROC_CNN}
    \end{subfigure}
    \caption{ViT-v2 Training and Validation Loss \& Accuracy}
    \label{fig:Vit-v2_curve}
\end{figure}

The AUC curve of the ViT models \ref{fig:ROC_ViT} shows a promising result, than compared to CNN AUC \ref{fig:ROC_CNN} curve.

The ROC curve for the "No Finding" class shows an AUC of 0.50, indicating the model performs as well as random chance when trying to differentiate healthy cases from potential diseases, highlighting the difficulty in identifying subtle differences between normal and abnormal findings. In contrast, the model performs somewhat better for diseases like Pleural Thickening (AUC = 0.53), Atelectasis (AUC = 0.51), and Pneumothorax (AUC = 0.51), where clearer visual signs may aid more accurate detection. However, it faces challenges with classes such as Cardiomegaly (AUC = 0.48), Emphysema (AUC = 0.52), and Hernia (AUC = 0.43). These challenges are primarily due to a data imbalance in our dataset, which results in limited availability of examples for certain conditions. 

\subsection{Attention Maps}
\begin{figure}
    \centering
        \includegraphics[width=\textwidth]{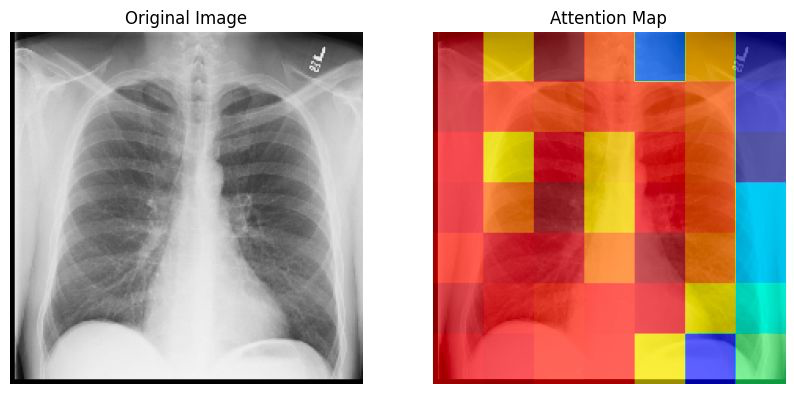}
    \caption{Attention map of a No Finding image}
    \label{fig:Attention map}
\end{figure}

To gain insights into the decision-making process of our Vision Transformer (ViT) model for chest X-ray image classification, we generated attention maps that highlight the regions the model focused on when making predictions. \ref{fig:Attention map}

We accomplished this by fine-tuning an InceptionResNetV2 model pre-trained on ImageNet, adding a ViT encoder and classification head on top. During inference, we obtained the attention weights from the last multi-head attention layer, which represent the importance assigned to different patches of the input image. 

By averaging these weights across heads and patches, resizing to match the input image size, and superimposing the attention map as a heatmap onto the original image, we could visualize the areas that were most influential for the model's predictions. 

\subsection{Limitations}
In our current setup, we trained on a subset of 85,000 images rather than the full NIH Chest X-ray dataset comprising of 112,120 images. Using only a subset of the data may impact the representativeness of the training data and generalizability of our findings.

Furthermore, our study could have explored subdividing the diseases and optimizing the transformers to only classify a particular disease. Optimizing models to focus on the detection of specific diseases rather than multiclassification, may have led to improved performance.

\section{Conclusion}
In this study, we compared the performance of Convolutional Neural Networks (CNNs), Residual Networks (ResNet), and Vision Transformers (ViTs) for multi-label classification of chest cancerous cells using the NIH Chest X-ray dataset. Our findings highlight that while CNNs and ResNet provide robust results, ViTs excel in diagnostic accuracy when pre-trained on extensive datasets like ImageNet-21k. This advantage is attributed to their sophisticated feature extraction through attention mechanisms, suggesting their potential for enhancing diagnostic processes in medical imaging. Future directions include optimizing these models with larger, well-annotated datasets and exploring disease-specific versions to further improve their clinical efficacy and integration into healthcare systems, thereby supporting more precise and timely medical decision-making.

\bibliographystyle{IEEEtranN}
\bibliography{references}

\clearpage

\appendix
\label{appendix}


\section{Models Evaluation Metrics Curves}

\begin{figure}[H]
    \centering
    \begin{subfigure}[b]{0.65\textwidth}
        \centering
        \includegraphics[width=\textwidth]{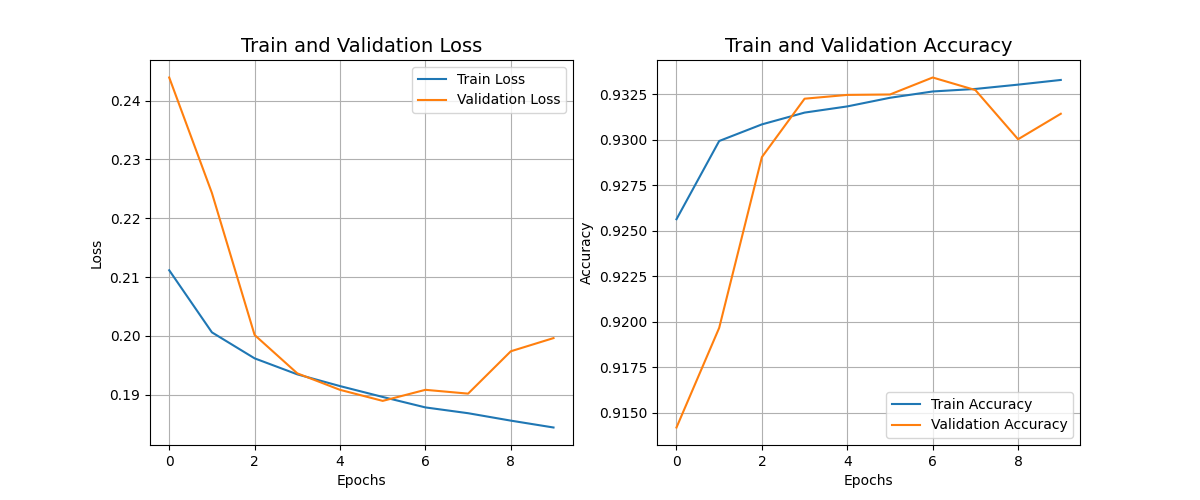}
        \caption{ResNet Training and Validation Loss \& Accuracy}
        \label{fig:ResNet_Curve}
    \end{subfigure}
    \hfill
    \begin{subfigure}[b]{0.65\textwidth}
        \centering
        \includegraphics[width=\textwidth]{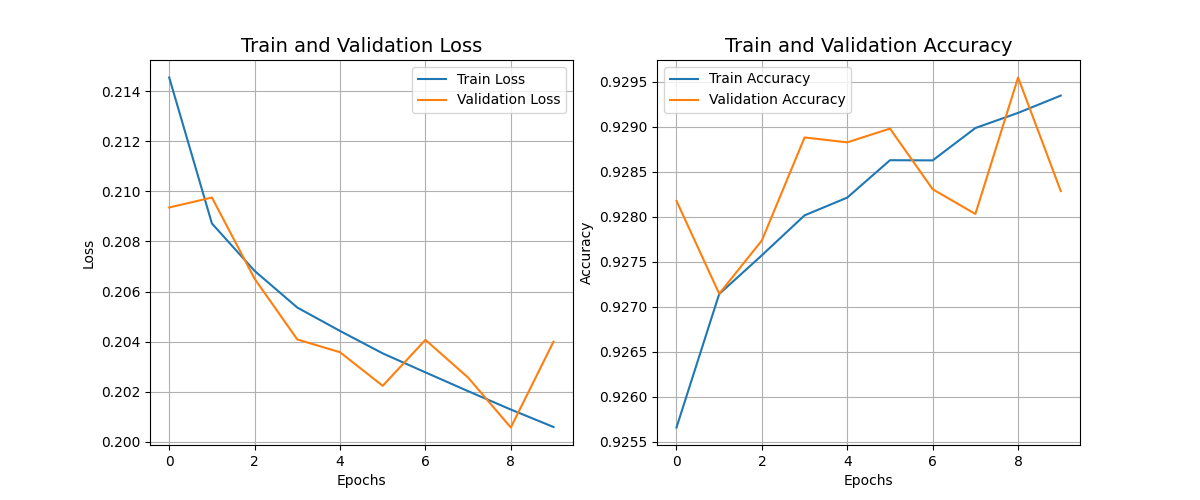}
        \caption{CNN Training and Validation Loss \& Accuracy}
        \label{fig:CNN_Curve}
    \end{subfigure}
    
    \vspace{0.5cm} 
    
    \begin{subfigure}[b]{0.45\textwidth}
        \centering
        \includegraphics[width=\textwidth]{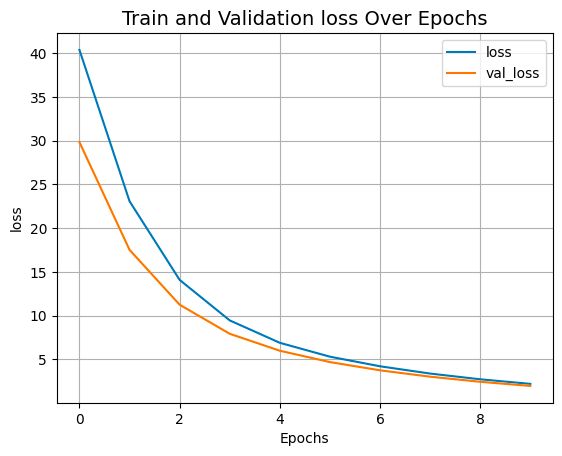}
        \caption{ViT-v2 Training Loss}
        \label{fig:training_loss}
    \end{subfigure}
    \hfill
    \begin{subfigure}[b]{0.45\textwidth}
        \centering
        \includegraphics[width=\textwidth]{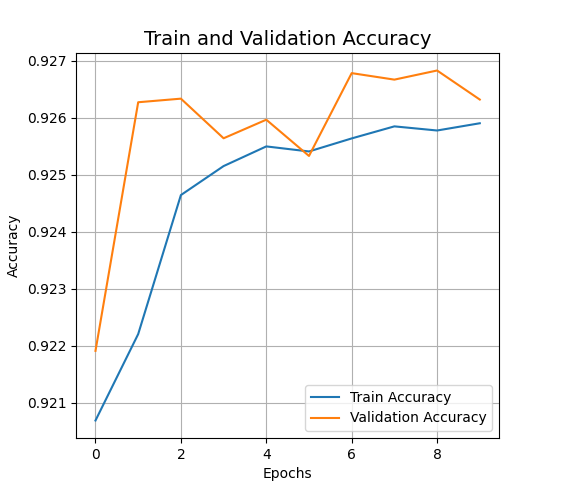}
        \caption{ViT-v2 Validation Accuracy}
        \label{fig:validation_accuracy}
    \end{subfigure}
    
    \vspace{0.5cm} 

    \begin{subfigure}[b]{0.9\textwidth}
        \centering
        \includegraphics[width=\textwidth]{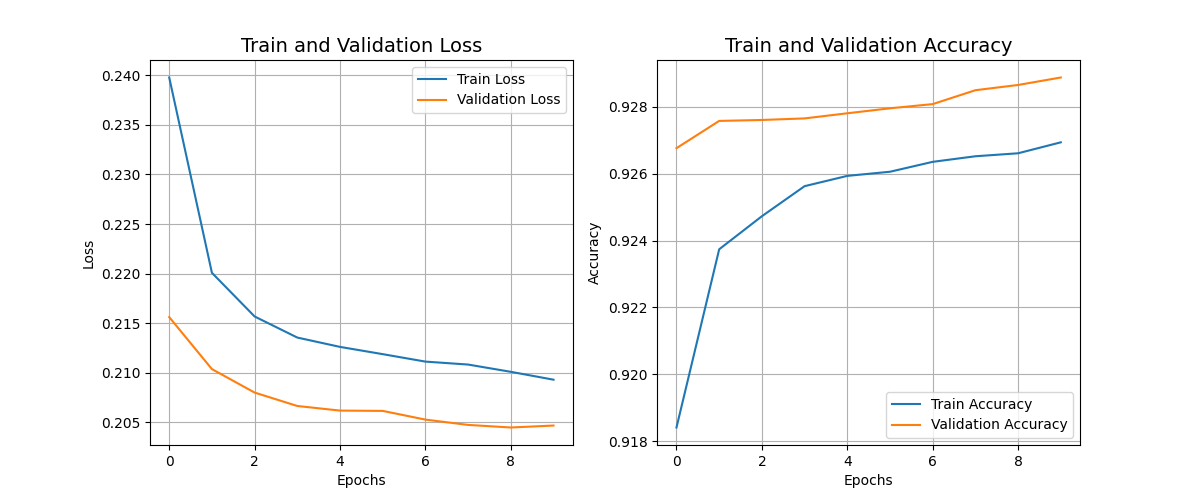}
        \caption{ViT-v1 Training and Validation Loss \& Accuracy}
        \label{fig:ViT-v1_Curve}
    \end{subfigure}

    \label{fig:combined_curves}
\end{figure}


\end{document}